\documentclass[%
reprint,
superscriptaddress,
 showpacs,preprintnumbers,
 amsmath,amssymb,
prb,
 floatfix,
]{revtex4-1}

\usepackage{booktabs}
\usepackage{array}
\newcolumntype{L}[1]{>{\raggedleft\arraybackslash}p{#1}}
\AtBeginDocument{
\heavyrulewidth=.08em
\lightrulewidth=.05em
\cmidrulewidth=.03em
\belowrulesep=.65ex
\belowbottomsep=0pt
\aboverulesep=.4ex
\abovetopsep=0pt
\cmidrulesep=\doublerulesep
\cmidrulekern=.5em
\defaultaddspace=.5em
}

\usepackage{subcaption}
\usepackage{soul}
\usepackage{hyperref}
\usepackage{color}
\usepackage{graphicx}
\usepackage{dcolumn}
\usepackage{bm}
\usepackage{cancel}
\usepackage[utf8]{inputenc}

\usepackage{upgreek}
\usepackage[normalem]{ulem}
\newcommand{\etal}{\textit{et al}.}
\newcommand{\ie}{i.e.}

\begin{document}

\title{Three-dimensional tomographic imaging of the magnetization vector field using Fourier transform holography}

\author{Marisel Di Pietro Martínez}
\email{Marisel.DiPietro@cpfs.mpg.de}
\affiliation{Université Grenoble Alpes, CNRS, Grenoble INP, SIMaP, 38000 Grenoble, France}
\author{Alexis Wartelle}
\affiliation{Université Grenoble Alpes, CNRS, Grenoble INP, SIMaP, 38000 Grenoble, France}
\affiliation{European Synchrotron Radiation Facility, F-38043 Grenoble, France}
\author{Carlos Herrero Martínez}
\affiliation{Université Grenoble Alpes, CNRS, Grenoble INP, SIMaP, 38000 Grenoble, France}
\author{Farid Fettar}
\author{Florent Blondelle}
\author{Jean-François Motte}
\affiliation{Université Grenoble Alpes, CNRS, Grenoble INP, Institut Néel, 38042 Grenoble, France}
\author{Claire Donnelly}
\affiliation{Max Planck Institute for Chemical Physics of Solids, Noethnitzer Str. 40, 01187 Dresden, Germany}
\author{Luke Turnbull}
\affiliation{Department of Physics, Durham University, Durham, DH1 3LE, UK}
\author{Feodor Ogrin}
\affiliation{School of Physics and Engineering, University of Exeter, Stocker Road, Exeter EX4 4QL, UK}
\author{Gerrit van der Laan}
\affiliation{Diamond Light Source, Harwell Science and Innovation Campus, Didcot OX11 0DE, UK}
\author{Horia Popescu}
\author{Nicolas Jaouen}
\affiliation{Synchrotron SOLEIL, Saint Aubin, BP 48, 91192 Gif-sur-Yvette, France}
\author{Flora Yakhou-Harris}
\affiliation{European Synchrotron Radiation Facility, F-38043 Grenoble, France}
\author{Guillaume Beutier}
\affiliation{Université Grenoble Alpes, CNRS, Grenoble INP, SIMaP, 38000 Grenoble, France}

\date{\today}

\begin{abstract}
In recent years, interest in expanding from 2D to 3D systems has grown in the magnetism community, from exploring new geometries to broadening the knowledge on the magnetic textures present in thick samples, and with this arise the need for new characterization techniques, in particular tomographic imaging.
Here, we present a new tomographic technique based on Fourier transform holography, a lensless imaging technique that uses a known reference in the sample to retrieve the object of interest from its diffraction pattern in one single step of calculation, overcoming the phase problem inherent to reciprocal-space-based techniques.
Moreover, by exploiting the phase contrast instead of the absorption contrast, thicker samples can be investigated.
We obtain a 3D full-vectorial image of a $800$\,nm-thick extended Fe/Gd multilayer in a $5\,\upmu$m-diameter circular field of view with a resolution of approximately $80$\,nm. The 3D image reveals worm-like domains with magnetization pointing mostly out of plane near the surface of the sample but that falls in-plane near the substrate.
Since the FTH setup is fairly simple, it allows modifying the sample environment. 
Therefore, this technique could enable in particular a 3D view of the magnetic configuration's response to an external magnetic field.
\end{abstract}

\maketitle

\section{Introduction}
\label{sec:intro}

Three-dimensional magnetic textures have recently attracted increasing interest both from fundamental and a technological point of view ~\cite{Streubel2016,fernandez2017,fischer2020,nguyen2015,may2019,kent2021creation,meng2021fabrication,donnelly2021experimental,donnelly2022complex}. This emergent field of research comes hand in hand with the need for new characterization techniques, in particular to obtain tomographic images of the magnetic textures. Among the wide variaty of magnetic microscopies, 
transmission based techniques offer the possibility to extend their capabilities to 3D, that is, to probe the magnetization as a vector field through the depths of the material.
Such capability has been demonstrated for neutrons~\cite{manke2010,hilger2018}, 
x-rays~\cite{Streubel2016,donnelly2017three} and electrons~\cite{wolf2019holographic,wolf2022unveiling}, at distinct length scales. 
The development done with neutrons allowed to image the magnetic domain distribution in the bulk, electrons permitted the characterization of the domain walls and observation of skyrmion tubes
in objects of approximately $100$\,nm thickness, whereas x-ray magnetic tomography allowed to observe new textures, such as Bloch points~\cite{donnelly2017three}, merons~\cite{hierro2020} and vortex rings~\cite{donnelly2020}, in samples from $200$\,nm thickness for soft x-rays up to $5\,\upmu$m using hard x-rays.

In particular, x-rays offer a range of microscopic and tomographic techniques 
well suited to the study of micron size samples with nanoscale resolution. 
The magnetic sensitivity is usually obtained by exploiting x-ray magnetic circular dichroism~\cite{van2014x}, \ie{}, an absorption contrast for opposite helicities of circular polarizations of the incident light.
High-resolution 2D imaging is routinely achieved with x-ray microscopes based on zone-plate optics. When operated in imaging mode (transmission x-ray microscopy, TXM), the resolution is given by the outermost zone width of the zone plate, whereas for scanning mode (scanning transmission x-ray microscopy, STXM) it is possible to achieve a higher resolution since the latter is given by the beam size~\cite{Fischer2006}. Both full-field TXM and STXM have been successfully extended into magnetic tomography techniques~\cite{Suzuki2018,witte2020,donnelly2022complex,hierro2020,hermosa2022}.

Exploiting the coherence of the beam can in principle provide a higher resolution, but more interesting is that it provides a phase contrast in addition to the absorption contrast, which shall be referred to here as x-ray magnetic circular birefringence.
This aspect is particularly appealing to investigate thick samples, since the magnetic phase contrast can remain sizable a few {eV} away from the absorption edge~\cite{Donnelly2016}, which in turn reduces the sample damage.
Coherence-based imaging techniques~\cite{Miao2015}, such as coherent diffraction imaging (CDI), Fourier transform holography (FTH)~\cite{stroke1965lensless} and ptychography, are 
well-suited to obtain 3D structural images~\cite{Chapman2006,guehrs2012soft,holler2017} and 2D magnetic images with nanometric resolution~\cite{Turner2011,eisebitt2004lensless,Tripathi2011}.
However, among the latter three techniques, only ptychography has so far been adapted to obtain full tomographic magnetic images~\cite{donnelly2017three,donnelly2020time,rana2021}.
Here we propose to extend FTH capabilities to 3D magnetic imaging.

\begin{figure*}[!t]
\begin{center}
    \includegraphics[width=0.98\linewidth]{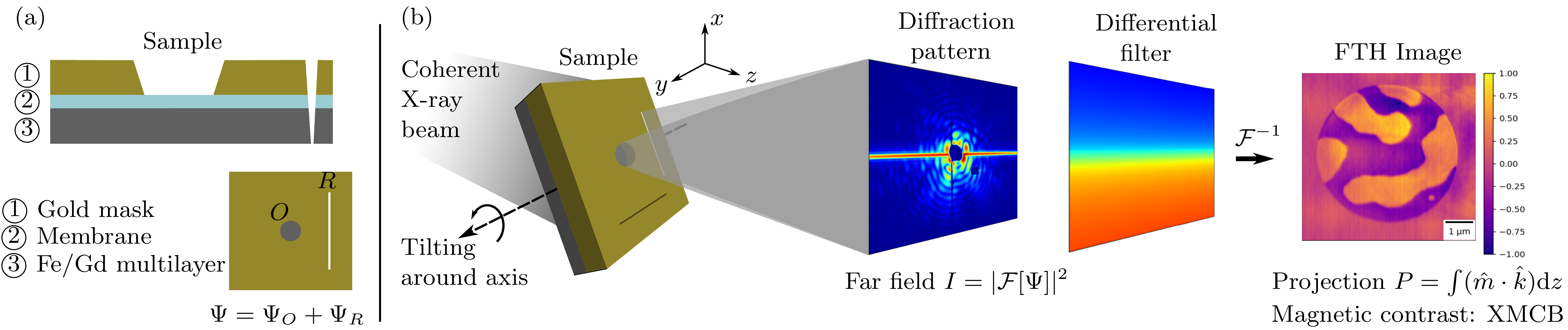}
\end{center}
\caption{Magnetic nanotomography based on Fourier transform holography (FTH): (a) The sample for FTH has three layers -- a gold mask, the membrane and the magnetic material of interest. (b) A circular window of $5\upmu$m-diameter is milled into the (opaque) gold layer, which coats the magnetic sample, to allow the x-rays to pass through. Two reference slits are also milled across the sample and coating. The coherent x-ray beam illuminates the whole sample. The complex x-ray amplitude after the sample, \ie{}, the exit wave, is denoted as $\Psi$. After filtering the reference from the diffraction pattern and applying an inverse Fourier transform, the magnetic projection is recovered. The magnetic contrast is obtained by x-ray magnetic circular birefringence (XMCB). Tilting the sample around axes $x$ and $y$ allows to probe all three components of the magnetization.}
\label{fig:fth}
\end{figure*}

The main asset of Fourier Transform Holography~\cite{stroke1965lensless} is 
being able to retrieve an image of the structure from the experimental data in only one deterministic step.
Moreover, it only requires a simple instrumental setup consisting of a pinhole to impose the high coherence of the incident beam, a rotating sample stage to select the magnetic projection and a beamstop -- protecting the high resolution 2D detector in the far-field of the sample~\cite{popescu2019comet}, which leaves space to implement the modification of the sample environment, such as controlling the temperature or applying an \textit{in situ} magnetic field.

Indeed, the complexity resides mostly in the sample preparation. The required sample consists of the object of interest $O$ and a known reference $R$ (described in terms of 2D, complex transmission functions), which interfere in the coherent beam (see Fig.~\ref{fig:fth}(a)). 
The holographic reconstruction provides an image which consists of the convolution of the object $O$ and the reference $R$. As a consequence, the resolution of FTH is limited by the reference size and quality. 
Additionally, phase retrieval algorithms can be used as a complementary method to improve the FTH resolution~\cite{flewett2012holographically}.

For extended references, following the HERALDO approach~\cite{guizar2007holography}, a linear differential operator specific to the chosen reference can be exactly calculated and consecutively applied to the measured intensity. In this way, the real-space image is deconvoluted with the reference, so that
a complex-valued image of the object can be retrieved 
in a single deterministic step (see Fig.~\ref{fig:fth}(b)), rather than following an iterative approach.
This image is equivalent to the object complex transmission coefficient 
if the object and the reference do not overlap~\cite{guizar2007holography}.

FTH has shown to be useful to obtain 2D images of the magnetization in flat samples \cite{duckworth2011magnetic,duckworth2013holographic,turnbull2020tilted,birch2020real}. 
Its inherent mechanical stability thanks to the integration of the reference in the sample itself makes FTH particularly interesting for time-resolved measurements~\cite{wang2012femto,bukin2016time}.
In fact, what is measured in forward scattering is a projection of the magnetization, just as with any other transmission technique~\cite{van2008soft}.
This is the component of $\hat{m}$ that is parallel to the beam direction $\hat{k}$ integrated through the material along the said direction $r_{k}$: 
\begin{equation}
P_{\hat{k}} = \int (\hat{m}\cdot\hat{k}) \mathrm{d}r_k.
\end{equation}

So whereas the first report of FTH focused on imaging the out-of-plane magnetization, \ie{}, the component perpendicular to the surface of the sample~\cite{eisebitt2004lensless}, if the sample is tilted the method also allows us to probe the in-plane magnetization components, using either a tilted reference hole~\cite{tieg2010imaging} or an extended reference~\cite{duckworth2013holographic}. Furthermore, it has also been shown that it is possible to use FTH to perform tomography and obtain the 3D electronic density~\cite{guehrs2012soft,guehrs2015mask}.

In this work, we go further and use 
FTH as a 3D full-vectorial magnetic imaging technique. 
To this end, we tilt the sample around two orthogonal axes perpendicular to the beam direction and, for each tilt, we measure a dichroic projection image. Acquiring a dual set of dichroic projections has been proven using other techniques to be sufficient to reconstruct not only the charge density of an object but also all three components of the magnetization in an entire three-dimensional structure\cite{donnelly2017three,donnelly2018tomographic,hierro2020}, including the inner configuration.

This paper is structured as follows.
In Sec.~\ref{sec:methods}, we describe the sample used to test the proposed technique, the experimental setup and we give the details regarding the data analysis.
In Sec.~\ref{sec:val}, we present a numerical validation of the method and analyze its limitations, followed by the experimental proof in Sec.~\ref{sec:results}. Finally, in Sec.~\ref{sec:conclusions}, we summarize the conclusions of this work. 

\section{Experimental details}
\label{sec:methods}

\begin{figure}[!b]
  \begin{center}
  \includegraphics[width=0.98\linewidth]{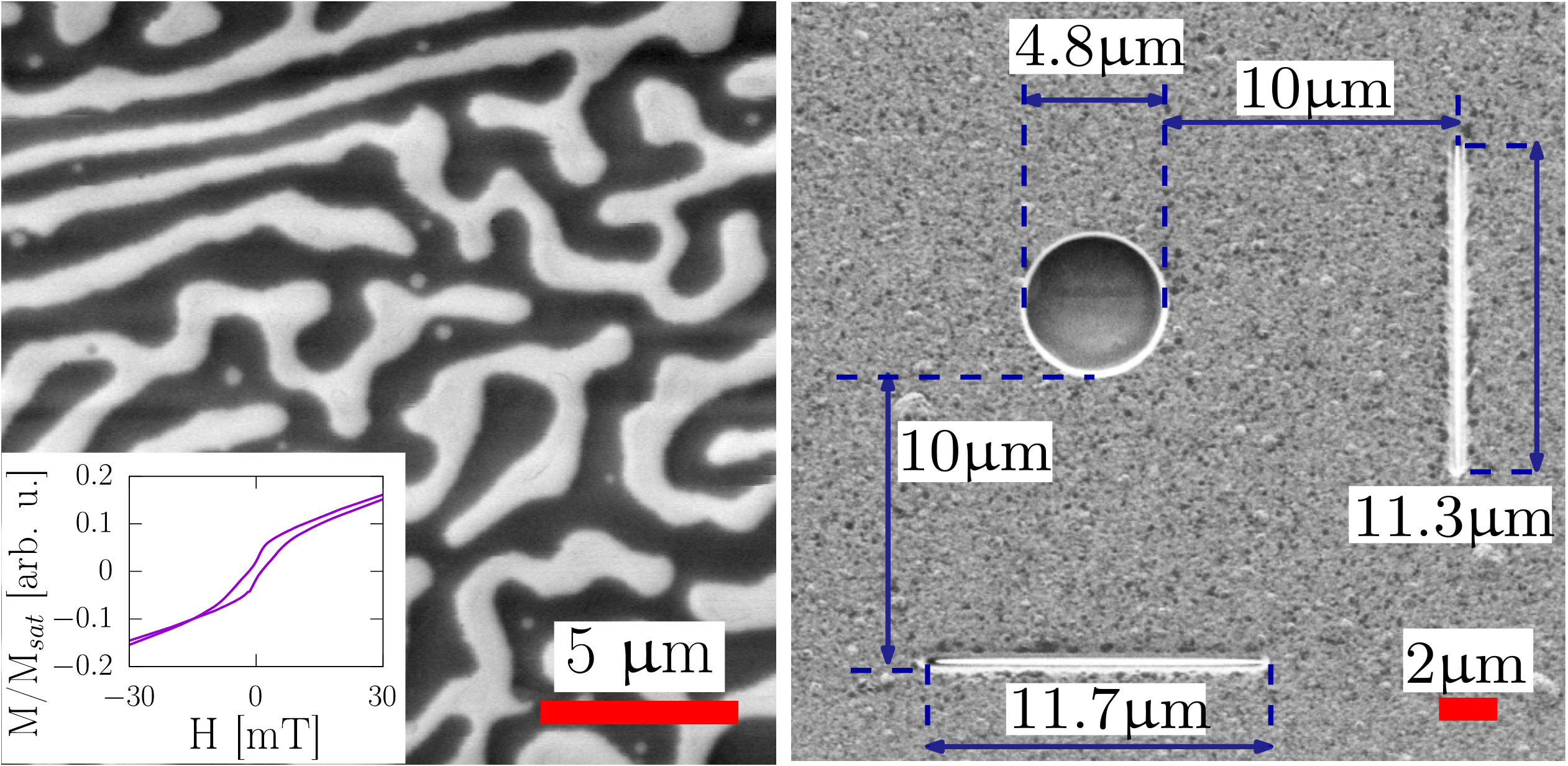}
  \end{center}
  \caption{Details of the sample for 3D Fourier transform holography: (a) MFM image of the Fe/Gd multilayer grown on one side of a Si$_3$N$_4$ membrane. Inset: In-plane magnetic hysteresis loop. (b) SEM image of the Au mask grown on the other side of the membrane. A circular aperture and two reference slits are milled for doing 3D-FTH tomography.}
  \label{fig:mfm-sem}
\end{figure}

We test the proposed tomographic method experimentally on an Fe/Gd multilayer which displays worm-like magnetic domains with a typical width of $1$\,$\upmu$m, as seen by magnetic force microscopy (MFM)\footnote{MFM images were acquired at room temperature (RT) and zero field with a low-moment PPP-LM-MFMR tip from Nanosensors, monitoring the resonance frequency shift during the 2nd pass at a lift height of $10$\,nm.} (Fig.~\ref{fig:mfm-sem}(a)). 
In-plane magnetization curves also show that, in spite of the dominating perpendicular magnetic anisotropy, it is expected to also have a persistent magnetic remanence as shown in the inset of Fig.~\ref{fig:mfm-sem}(a).
This coexistence of both in-plane and out-of-plane magnetization promises an intriguing 3D configuration, that cannot be mapped by 2D imaging techniques.

The multilayer was sputtered at room temperature with deposition rates reaching $1.7$, $0.5$ and $1.2$\,\AA/sec for Ta, Fe and Gd, respectively, and limit pressure $7.10^{-8}$\,mbar.
The nominal stacking for this sample is Ta($6$)/[Fe($0.45$)/Gd($0.96$)]$_{600}$/Ta($6$) where the thicknesses are expressed in nm and $600$ is the number of repetitions of the bilayer. The average composition of this sample 
was measured with energy-dispersive x-ray spectroscopy (EDX) to be Fe$_{0.667}$Gd$_{0.333}$ and the total stack thickness as determined from scanning electron microscopy (SEM) is approximately $800$\,nm.


The sample was grown on a $300$\,nm-thick Si$_3$N$_4$ membrane suitable for x-ray measurements. This membrane was covered with a $1700$\,nm-thick gold mask which is opaque to soft x-rays. The mask has also four $5$\,nm-thick Ti layers grown intercalated with the Au to prevent the formation of large Au grains and the subsequent leakage of x-rays.
Then we milled a circular aperture of diameter $d=5$\,$\upmu$m into the gold mask using focused ion beam (FIB) to allow the transmission of x-rays. This aperture represents the object $O$ in the FTH approach (Fig.~\ref{fig:fth}). To create the references $R$, two thin slits of 
length of $11\,\upmu$m, perpendicular to each other and at a distance $10\,\upmu$m of the circular aperture, were milled across the coating and the sample (Fig.~\ref{fig:mfm-sem}(b)). The location and length of the slits meet the HERALDO separation conditions, which prevent the overlapping of the deconvoluted object and reference images~\cite{guizar2007holography}.
The width of the slits, which goes down to $\sim80$\,nm across the 2.8\,$\mu$m total thickness of the full stack (see transversal slice of the slit in Fig.~\ref{fig:fth}(a)), limits the resolution in one of the transverse directions in individual 2D images, while the resolution in the other transverse direction is limited to  $\sim50$\,nm by the sharpness of the slit end.

The FTH data presented in this work was mainly acquired on the COMET endstation~\cite{popescu2019comet} at SEXTANTS beamline of SOLEIL synchrotron, and complementary data was measured at beamline ID32 of the ESRF. 
Both beamlines use similar setups.
Circularly polarized x-rays are delivered by a helical undulator and the energy of the beam tuned by a grating monochromator.
The coherence of the beam is ensured by a set of apertures in front of the endstation.
The small angle coherent diffraction patterns are acquired
on an area detector with a CCD camera (SEXTANTS), or a CMOS camera~\cite{desjardins2020backside} (ID32).
The geometrical settings were such that the pixel size of the direct space images was $25$\,nm at SEXTANTS and $50$\,nm at ID32.
To allow for tilting along two orthogonal axes, an azimuthal rotation of the sample holder was implemented, in addition to the existing tilt rotation. This setup is also compatible with the laminography geometry~\cite{witte2020}.

To acquire the required dual set of projections, we tilt the sample around two orthogonal axes corresponding to the directions of both slits. By tilting around the $x$ axis according to Fig.~\ref{fig:fth}, for example, the vertical slit is shadowed by the thickness of the sample, while the horizontal slit is not, hence the latter serves as the holographic reference for the measurements. In the same way, by tilting around the $y$ axis, the horizontal slit is now obscured and the vertical slit serves as the reference. 
Only close to normal incidence can both slits serve as a reference. 
We measured projections for $22$ tilt angles in total, getting $3$ images per polarization for each, with a total acquisition time of $130$\,ms per image.
The FTH measurements were performed at room temperature and at remanence.

The FTH images were reconstructed using a Python notebook based on the one provided in Ref.~\cite{birch2020real}, which in turn follows the HERALDO method~\cite{guizar2007holography}. 
The FTH reconstruction algorithm provides complex-valued images, from which we extracted the phase since this quantity is proportional to the projection of the magnetization.
To maximize the magnetic contrast of the images, we worked at $704.6$ eV, which is $2.1$ eV below the Fe $L_3$-edge.
See the Appendix for more details on this.

Once all the measurements are processed and the set of projections is obtained, they are used as input for reconstructing the 3D magnetic configuration.
To that end, we developed the PyCUDA library \verb|magtopy|~\cite{magtopy}.

The reconstruction algorithm is based on the gradient descent method which has already been shown to be able to successfully reconstruct full-vectorial 3D magnetization configurations~\cite{donnelly2018tomographic}. Starting with an initial guess for the 3D magnetic structure $\hat{m}_0(x,y,z)=\vec{0}$, the next update is directed by minimizing the error metric
\begin{equation}
 \epsilon = \sum_\phi \sum_{x,y} \left(P^{(\phi)}(x,y) - P_{m}^{(\phi)}(x,y)\right)^2,
\end{equation}
where $\{P_{m}^{(\phi)}\}$ is the measured set of projections and $\{P^{(\phi)}\}$ is the one calculated from the guess as 
\begin{equation}
 P^{(\phi)}(x,y) = \sum_z R^{(\phi)}[\hat{m}] \cdot \hat{z}.   
\end{equation}
For each tilt angle $\phi$, the rotation matrix $R^{(\phi)}$ is applied to $\hat{m}$.
Once the gradient $\frac{\partial \epsilon}{\partial \hat{m}}$ is calculated, the structure is updated according to 
\begin{equation}
\hat{m}_{new} = \hat{m} - \alpha \frac{\partial \epsilon}{\partial \hat{m}}.
\end{equation}
We included a step optimization according to which the best step $\alpha$ is estimated by imposing the condition
\begin{equation}
\epsilon_{k - 1} - \epsilon_k > v,
\end{equation}
so that the error decreases sufficiently in each step, that is, more than a certain value $v$.

It is worth noting that the chosen programming language for the library, PyCUDA, provides the interoperability of Python while taking advantage of high performance computing.
The main algorithm is capable of reconstructing the 3D magnetic configuration of a $200^3$ voxels cube, or a ($5\,\upmu$m)$^3$ cube  considering a pixel size of $25$\,nm, in one minute~\footnote{We measured a speed-up of $\times75$ in a GeForce RTX 3090 compared to the serial version of the code run in a CPU.}.

\section{Validation of the reconstruction algorithm}
\label{sec:val}
 

To validate the vectorial reconstruction of the magnetic configuration,
as well as to understand its limitations,
we considered one of the most relevant and common problems that can arise during the experiment, which is having a reduced angular range for the tomography, also known as missing wedge and addressed in Ref.~\cite{hierro20183d} with a different reconstruction algorithm.
Another problem of FTH is the artifacts related to the imperfection of the references, for example an irregular slit end causes the overlay of weaker replicas over the main reconstruction.
However, since these artifacts are bound to the sample fabrication stage,
we will not address these in the present discussion.

\begin{figure}[!bt]
\begin{center}
\includegraphics[width=0.98\linewidth]{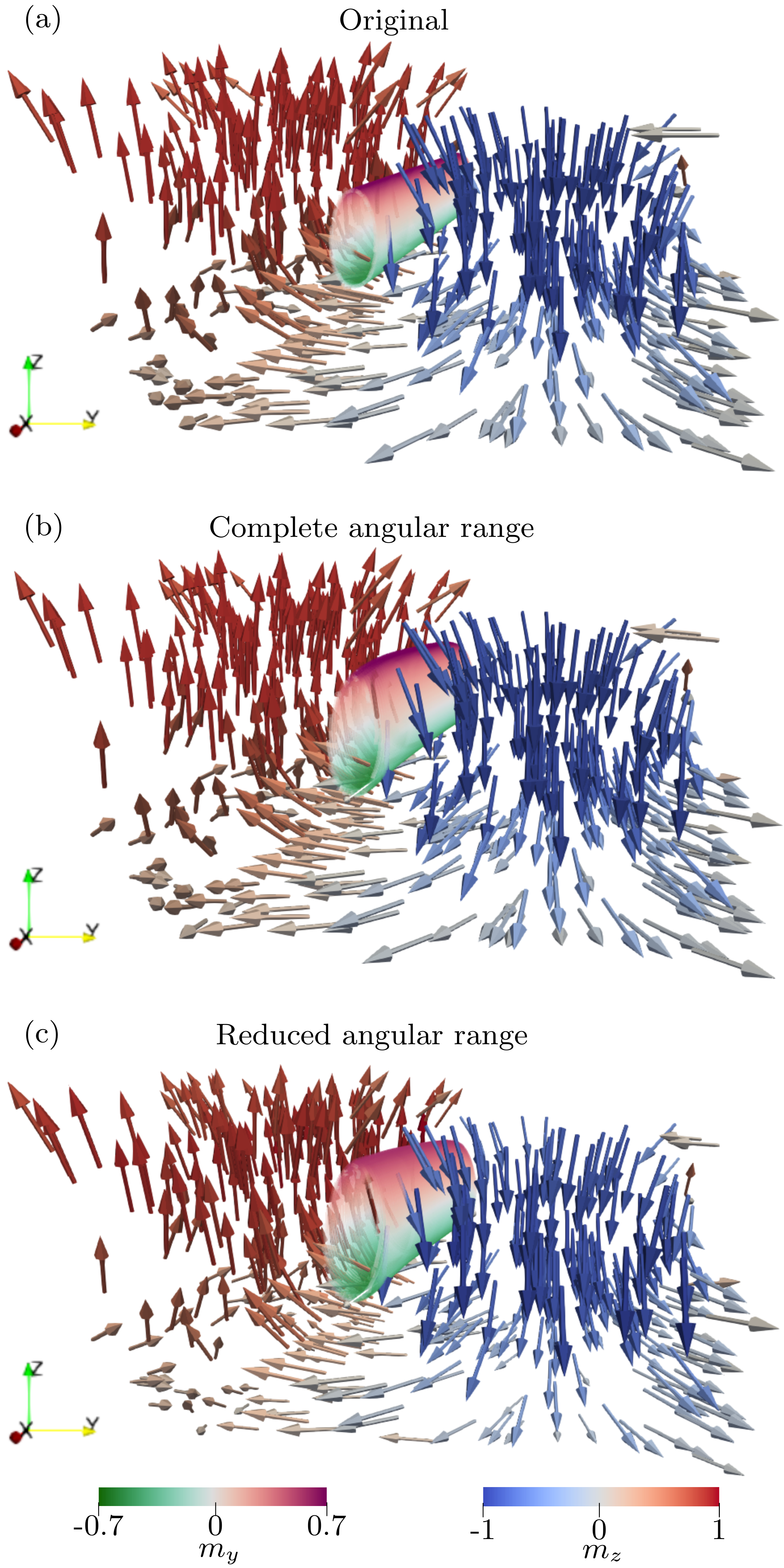}
\caption{Comparison between the (a) original magnetic configuration from simulations and the reconstructed configurations from (b) a complete angular range which includes projections of the sample tilted from $-90$\textdegree{} to $90$\textdegree{}, and (c) the reduced angular range which only includes tilts from $-45$\textdegree{} to $45$\textdegree{}. Streamlines in the center of the structure highlight the core of the domain wall. The normalized reduced mean square error in each case is: (b) NRMSE$(m_x) = 3\%$, NRMSE$(m_y) = 2\%$, NRMSE$(m_z) = 2\%$ and (c) NRMSE$(m_x) = 24\%$, NRMSE$(m_y) = 5\%$, NRMSE$(m_z) = 5\%$.}
\label{fig:sim}
\end{center}
\end{figure}

We used as a test case the simulated magnetic configuration from Ref.~\cite{beutier2005characterization} which has a size comparable to the experimental sample described above. The configuration, displayed in Fig.~\ref{fig:sim}(a), shows two main domains with opposite out-of-plane (z-axis) magnetization and a domain wall with a Bloch-type core and two opposite Néel closure caps. The streamlines shown in the center of the structure highlight the position of the Bloch core.


Measuring several projections tilting the sample around $180$\textdegree{} leads to highly accurate reconstructions of the magnetic configuration, as can be seen in Fig.~\ref{fig:sim}(b). There we can observe only a slight deformation of the streamlines in the domain wall core. The normalized reduced mean squared error (NRMSE
\footnote{The normalized reduced mean squared error is defined as follows, \begin{equation*} \mbox{NRMSE} = \frac{1}{X_{\mbox{max}} - X_{\mbox{min}}} \left[ \frac{1}{N} \sum_{i=1}^N (X_i - X_{0,i})^2 \right]^{\frac{1}{2}}, \end{equation*} where $X$ and $X_0$ are the reconstructed and the original 3D configuration, respectively, and $N$ is the size of the structure.})
calculated in this case is smaller than $3$\% for all three components of the magnetization.

However, the accessible angular range is usually limited experimentally, for instance by the geometrical constraints of the setup, and in particular by the geometry of the supporting membrane and its frame, which may shadow the object of interest at shallow incidence angles.
Therefore, we simulated projections for tilting angles ranging from $-45$\textdegree{} to $45$\textdegree{} to match the accessible ones in the experiment. The magnetic configuration reconstructed from the latter set is shown in Fig.~\ref{fig:sim}(c).

We observe that the NRMSE of one of the in-plane components, $m_x$, increases to $24$\%.
The increased error is mainly due to the missing wedge effect.
In particular, the magnetization along $x$ in this simulated system has a different behavior through the thickness of the sample than the rest, \ie{}, there is a larger component near the substrate that is not present near the surface. Compare the magnetic vectors on the top of the structure with the ones from the bottom: the former point mainly in the $z$ direction, while the latter are significantly tilted towards $x$.
The information on this inhomogeneity is lost when no projections are given between $45$\textdegree{} and $90$\textdegree{}. Nevertheless, the NRMSE of the other two components remains at $5$\%.
A similar effect has been observed also in simulated Py discs measured between $-55$\textdegree{} and $55$\textdegree{} and reconstructed with a different algorithm~\cite{hierro20183d}. From the streamlines in the center of the structure, we can see in detail how the walls are affected. In particular for the Néel caps, we see that $m_y$ is weaker compared to the original.

Altogether, note that the main features in the structure, namely the two opposite domains and the domain wall
are successfully recovered and fully recognizable, which grants the method a robustness against the angular limitation.

\section{Experimental results}
\label{sec:results}

\begin{figure}[!tb]
\begin{center}
\includegraphics[width=0.98\linewidth]{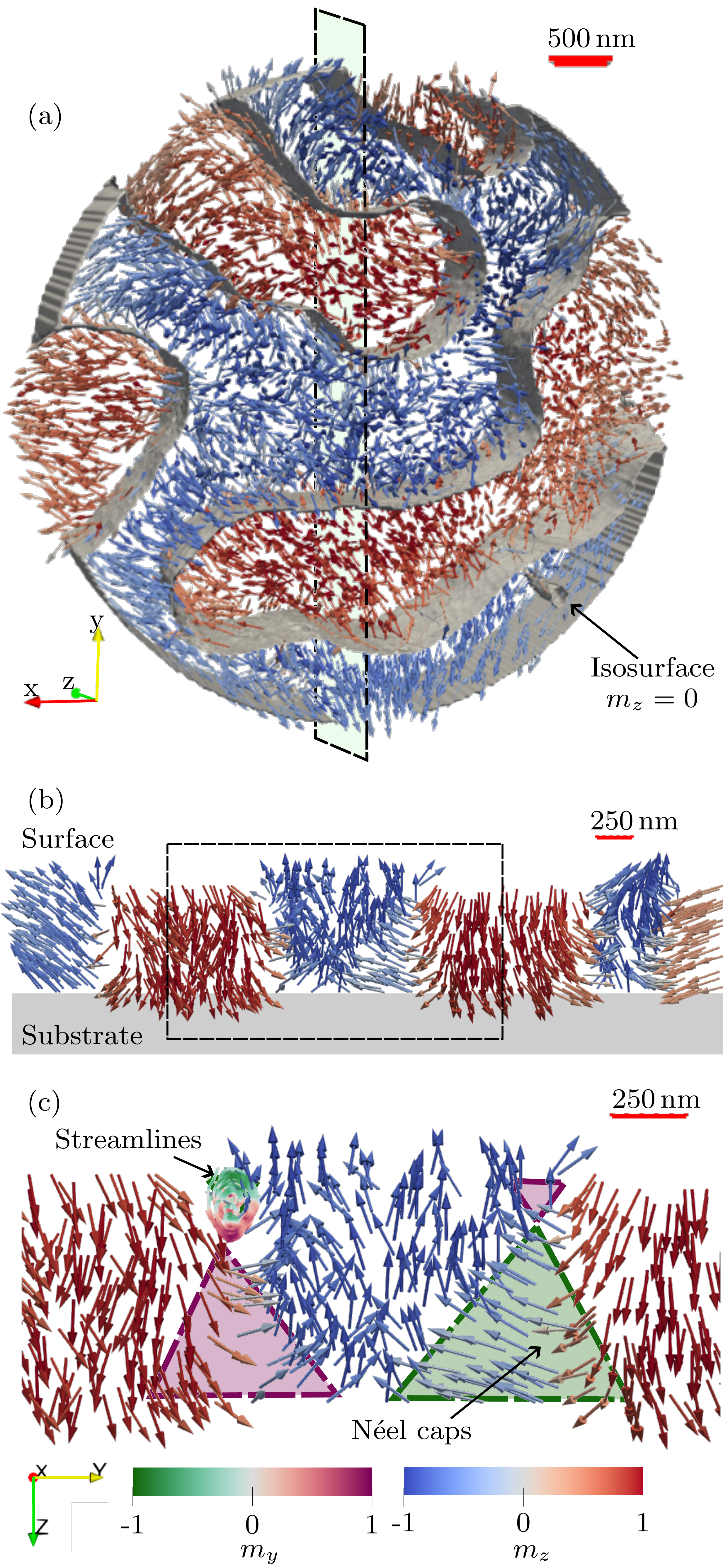}
\caption{3D magnetic image of the Fe/Gd multilayer obtained using Fourier transform holography with the two reference slit setup: (a) Overview of the magnetic vector field imaged through the circular aperture. In grey, the isosurface for $m_z=0$ hints mostly prismatic domains. The dashed rectangle indicates where the slice for (b) was taken, and in (b) the dashed rectangle indicates the area taken to show in more detail in (c). Both (b) and (c) have the same axes orientation.}
\label{fig:reconsabc}
\end{center}
\end{figure}

Now let us return to the experiments.
In Fig.~\ref{fig:reconsabc}(a), we show the full three-dimensional reconstruction of the magnetization vector field for the $800$\,nm-thick Fe/Gd multilayer described in Sec.~\ref{sec:methods}. Two kinds of domains appear: one with the magnetization pointing mainly towards the surface of the disk (in negative $z$ direction) and another with the magnetization pointing mainly away from it (in positive $z$ direction).
The general aspect of the magnetic structure is consistent with the MFM images performed on a full film (Fig.~\ref{fig:mfm-sem}(a)). More interesting is the depth structure, which 2D measurements cannot capture.

The isosurface for $m_z=0$ is also displayed as an overlay in Fig.~\ref{fig:reconsabc}(a) and it shows the location of the walls that separate the two domains. From this we can observe that the shape of the domains as seen from the surface spans through the thickness, so that the isosurface appears perpendicular to the surface. In another words, the volume of each domain has a prismatic shape. In particular, a small tube, possibly a skyrmion, can also be spotted in the lower-right corner of the structure.
Indeed, dipolar skyrmions have been reported in Fe/Gd multilayers~\cite{montoya2017tailoring} and seem to be present in the MFM measurement from Fig.~\ref{fig:mfm-sem}(a) as well.

In Fig.~\ref{fig:reconsabc}(b), we present a transversal slice along the $y$ axis, through the middle of the sample, to show in detail the magnetization vector field.
Here we observe
that the in-plane component of the magnetization increases close to the substrate. In that area we can distinguish the Néel caps. Close to the borders we can notice the magnetization vector falling into the $y$ direction. This is an artifact that comes from the missing information of the borders for an extended system and it affects mostly an outer ring of approximately $500$\,nm. This represents a limitation on the maximum field of view of the method used, that can be overcome by patterning a finite structure centered in the FTH aperture, as opposed to an extended one.

In Fig.~\ref{fig:reconsabc}(c), we show an area of the previous slice in more detail. Close to the top left corner, the streamlines help identifying the area of the Bloch core, similarly to the simulated system in Sec.~\ref{sec:val}. The color code of the streamlines highlights that the upper area have negative $m_y$ whereas the bottom have positive $m_y$, corresponding to the two Néel caps. The colored triangles correspond to the area where $m_y>0.5$, and they represent the Néel caps. It can be observed that the Néel caps closer to the substrate are larger than the ones close to the surface.
To estimate the width of the domain walls, we measured $m_z$ profile along $y$ in the first layer close to the surface ($z=0$\,nm) and in the last layer near the substrate ($z=775$\,nm), and we fitted a hyperbolic tangent. This will give us the domain wall width convoluted with the spatial resolution.
We obtained a width of $100$\,nm in the surface and $325$\,nm near the substrate. If we do the same for the position of the core marked by the streamlines in Fig.~\ref{fig:reconsabc}(c) ($z=200$\,nm), we obtain a width of $63$\,nm.



\begin{figure}[!htb]
  \begin{subfigure}{0.44\textwidth}
    \includegraphics[width=\linewidth]{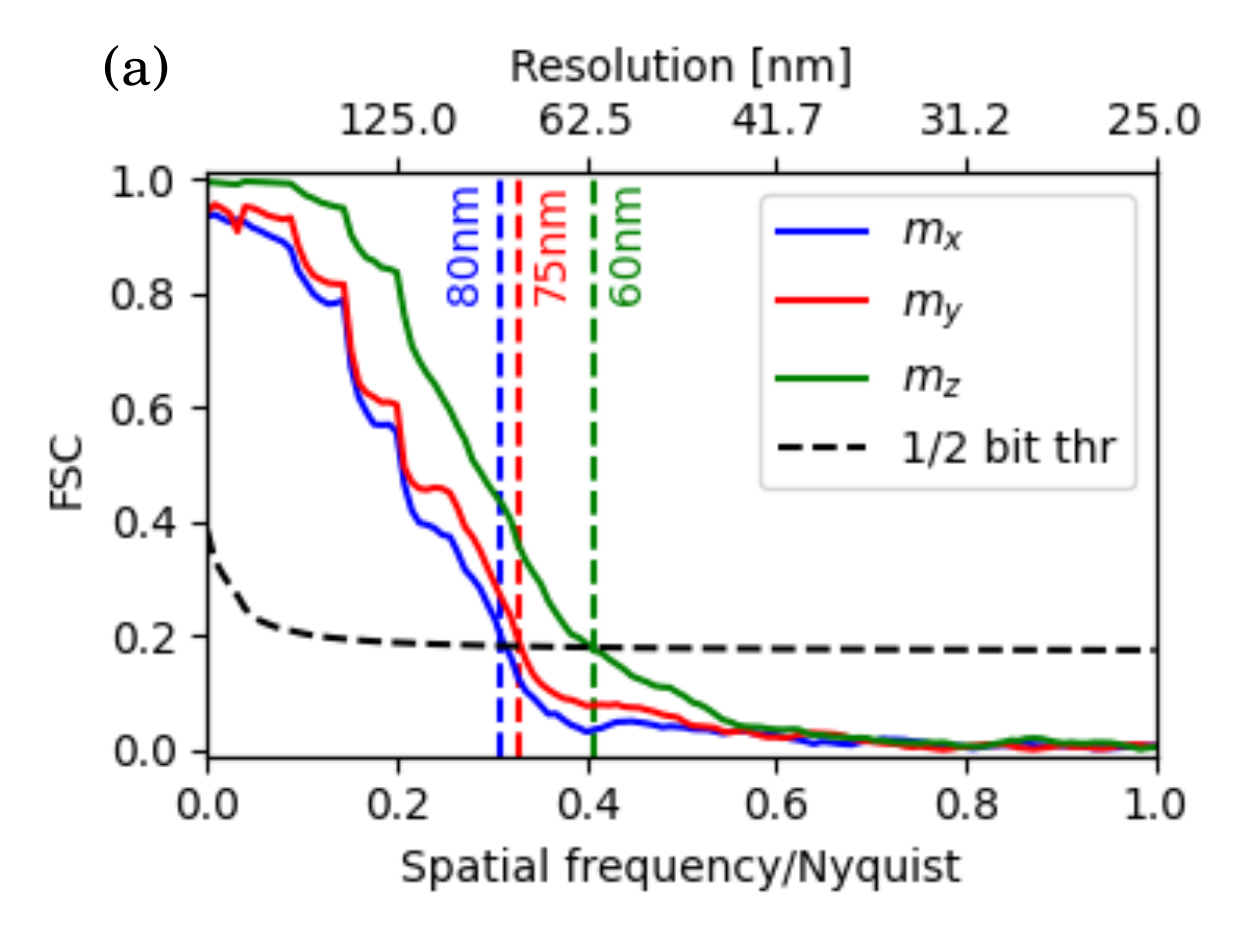}
  \end{subfigure}

\begin{subfigure}{0.5\textwidth}
    \includegraphics[width=\linewidth]{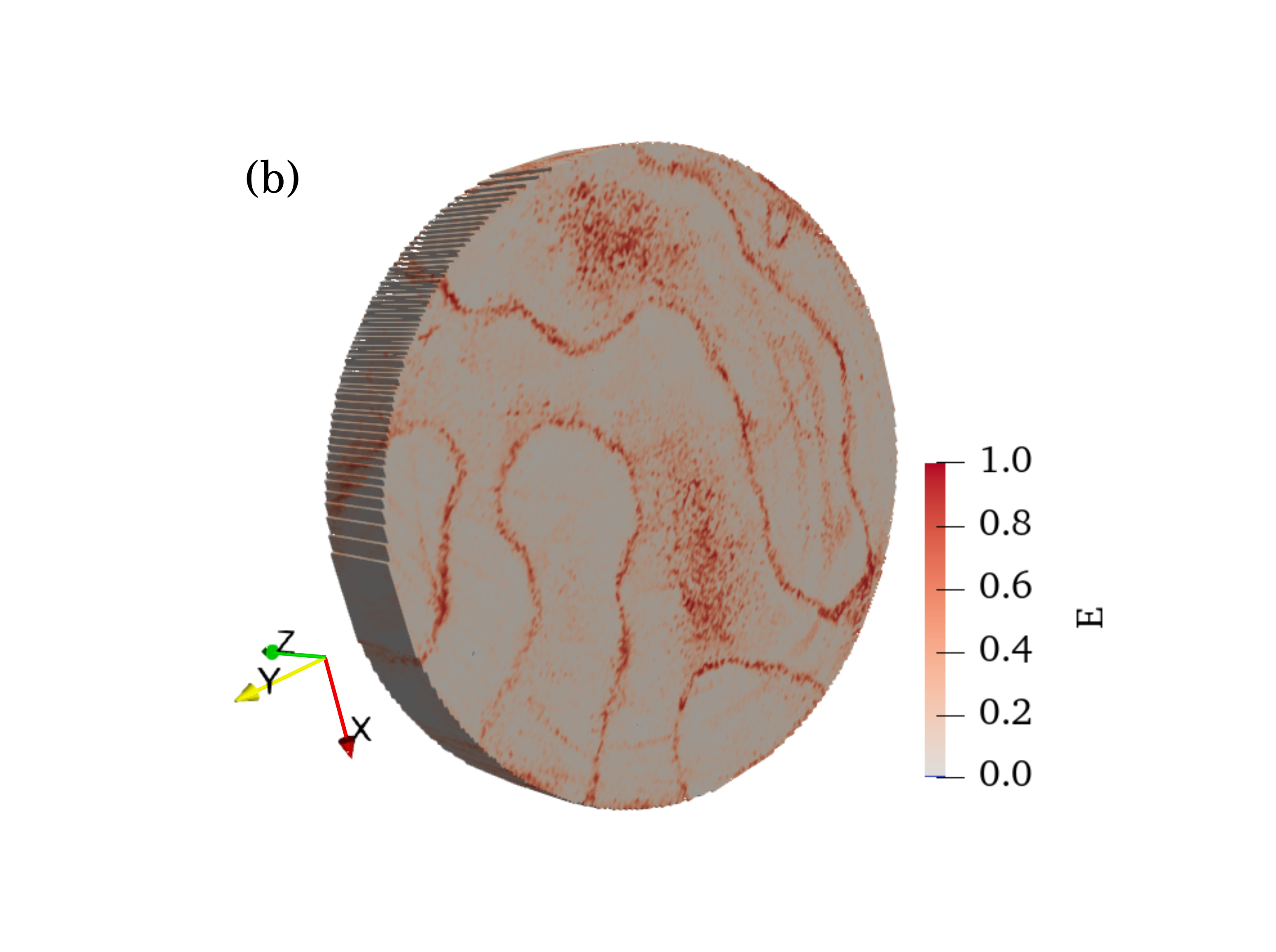}
    \end{subfigure}
\caption{Spatial resolution and error in the 3D magnetic image of the Fe/Gd multilayer: (a) Spatial resolution estimated via Fourier Shell Correlation (FSC) for the three magnetization components. The $1/2$-bit threshold (dashed line) is used to estimate these values. (b) Error $E$ between the reconstructions acquired with two complementary projection sets. Only $1$\% of the reconstruction has an error larger than 0.9 which means a complete opposite orientation of the magnetic moments. These are in turn concentrated in areas with artifacts coming from the FTH projection measurements and some specific regions of the domain walls.}
\label{fig:recons}
\end{figure}

The spatial resolution was estimated by calculating the Fourier shell correlation (FSC)~\cite{pynx} between two independent reconstructions. To this end, we split the projection set in two, and obtain a reconstruction configuration for each. We used the $1/2$-bit threshold criterion to ascertain the value of the spatial resolution\cite{van2005fourier}. We show these curves in Fig.~\ref{fig:recons}(a). The spatial resolution is $80$, $75$ and $60$\,nm for each of the components of the magnetization: $m_x$, $m_y$ and $m_z$, respectively.
These numbers are in-between the width of the slits ($\sim80$\,nm) and the sharpness of the slit ends, estimated to $\sim50$\,nm from individual 2D images. In particular, the resolution for $m_z$ matches the size of the Bloch core reported above.
2D FTH could achieve significantly higher resolution with a thinner and sharper reference ($17$\,nm claimed in Ref.~\cite{turnbull2020tilted}), which in turn would improve the resolution in the 3D reconstruction. 
For comparison, in previous soft x-rays dual-axis magnetic tomography based on transmission microscopy, a resolution of $85$\,nm for a $400$\,nm-thick film~\cite{hierro2020} and $10$\,nm for a $120$\,nm-thick superlattice~\cite{rana2021direct} has been reported.

While the FSC quantifies the resolution on average, to get a sense of the spatial localization of the error, we present also  Fig.~\ref{fig:recons}(b). Here, we show the error for each voxel of the reconstructed structure, defined as 
\begin{equation}
\begin{split}
&E = \\
&\frac{1}{2}\sqrt{(m_{x,1} - m_{x,2})^2 + (m_{y,1} - m_{y,2})^2 + (m_{z,1} - m_{z,2})^2},
\end{split}
\end{equation}
where the components with subscript $1$ and $2$ correspond to the two different projection sets used to calculate the FSC as described above. Note how the error is mainly concentrated approximately $75$\,nm around the area of the domain walls as well as in specific regions of the domains. The latter can be directly related to FTH measurement artifacts previously observed in the projection images, and in the reconstruction, these affect the inner layers (larger $z$) the most, doubling its value for the layer closest to the substrate.
Altogether, this shows that the 3D reconstruction concentrates its reliability in the domain area. 

\begin{figure}[!b]
\begin{center}
    \includegraphics[width=0.98\linewidth]{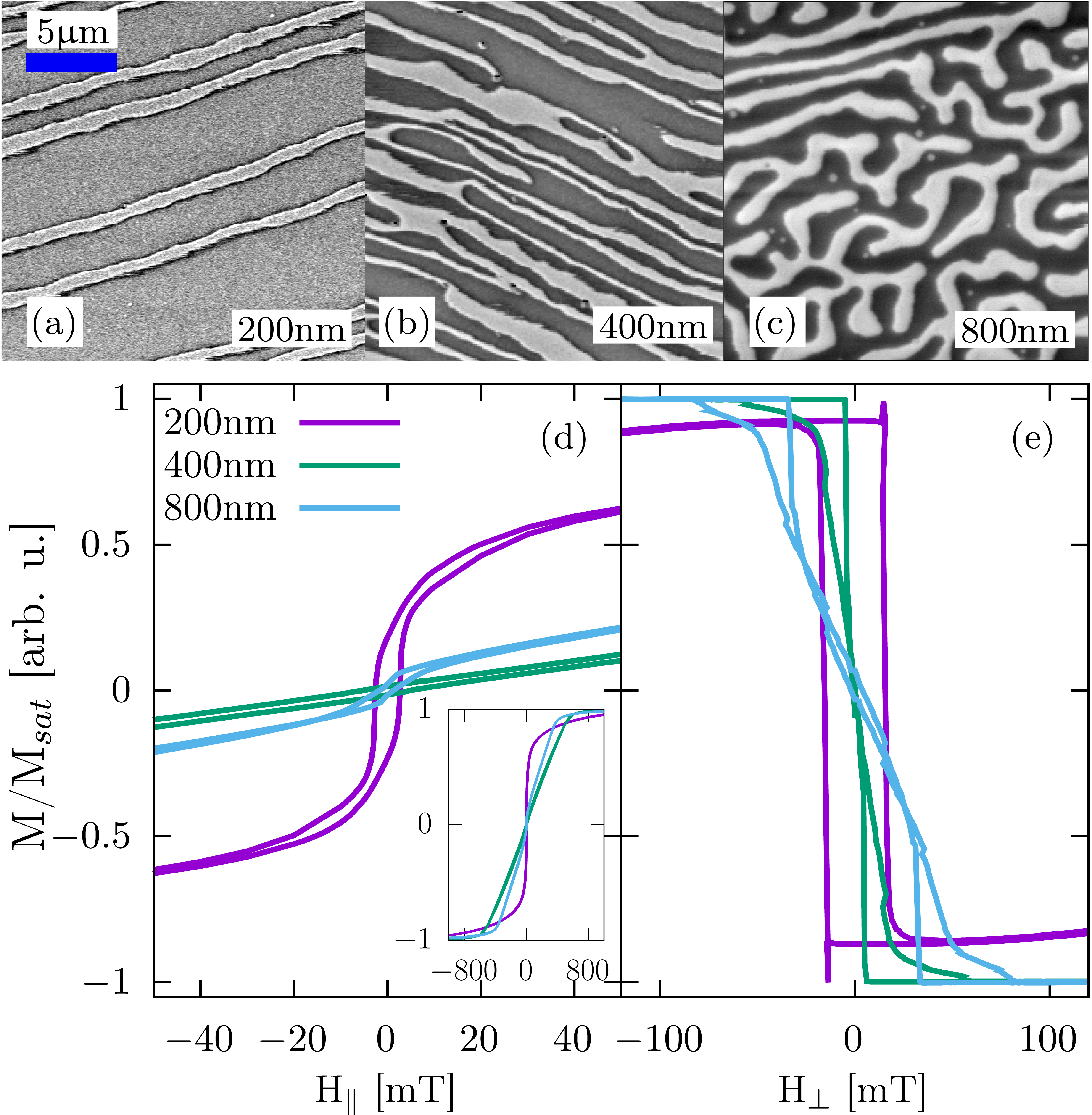}
    \caption{Magnetic behavior of Fe/Gd multilayers with different film thicknesses given by the total amount of bilayers stacked in each case: MFM images in remanence for $200$ (a), $400$ (b) and $800$\,nm (c) of Fe/Gd; in-plane (d) and out-of-plane (e) hysteresis loops for the same three samples.}
    \label{fig:squid}
\end{center}
\end{figure}

Magnetometry and MFM measurements can shed light on the observed phenomenon of in-plane magnetization predominance close to the substrate.
We considered two other samples that are thinner than the one used for 3D-FTH. While the sample used in the 3D-FTH experiment has $N=600$ repetitions of [Fe($0.45$\,nm)/Gd($0.96$\,nm)], the other two have $150$ and $300$ repetitions of the same Fe/Gd bilayer, corresponding to a thickness of $200$\,nm and $400$\,nm, respectively\footnote{The multilayers for magnetometry and MFM investigations were grown on a Si/SiO$_2$(1000\AA) substrate.}. The hysteresis loops of these three, as well as MFM images of the surface of each sample at remanence, are shown in Fig.~\ref{fig:squid}.

Extraordinary Hall effect (EHE) measurements provide sensitivity to the out-of-plane magnetization component.
The thinnest sample's ($200$\,nm) EHE data reveals a square hysteresis loop indicating perpendicular anisotropy (Fig.~\ref{fig:squid}(e)). With this in mind, we can understand its MFM image (Fig.~\ref{fig:squid}(a)). The broad domains correspond to mostly up and down magnetization. The large size is the reason for the contrast being stronger close to the domain walls since
a Néel character is expected close to the surface. This is expected for films that are thicker than the dipolar-exchange length like this one, since this allows for a better flux closure \cite{Hubert1998}.

The behavior of this $200$\,nm-thick sample is however more complicated than a typical perpendicular system with only uniaxial anisotropy. Indeed, the in-plane magnetometry data shown in Fig.~\ref{fig:squid}(d) features a clear hysteresis with sizeable remanence, as well as a large in-plane susceptibility up to significant values of $M/M_\mathrm{s}$. 
Both characteristics disappear at larger thicknesses \ie{}, for the $400$ and $800$\,nm-thick multilayers, which indicates less favorable in-plane magnetization. In turn, MFM images show that the large domains give way to perpendicular 
stripe-like domains and eventually worm domains when the thickness increases (see Fig.~\ref{fig:squid}(b) and (c)). 
When measuring thinner samples, the surface, to which the MFM is sensitive, is closer to the substrate, so we get more sensitive to any influence the substrate can have on the rest of the sample. 
Hence, this results hint at a decrease of the out-of-plane component 
near the substrate.

The Gd content has an specific effect on the sample behavior. 
Specifically, this sample effectively displays a ferrimagnetic behavior, such that the magnetic moments of Fe and Gd are antiferromagnetically coupled.
This is observed by EHE measurements (Fig.~\ref{fig:squid}.(d-e)) showing inverted loops.
Indeed, at our average sample composition, the corresponding alloy's magnetization is dominated by Gd~\cite{Hansen1989}, whereas the EHE is expected to be more sensitive to the perpendicular magnetization of Fe in this material~\cite{McGuire1976}, and since the Fe is antiferromagnetically aligned with the Gd, the magnetic loop is consequently expected to be inverted~\cite{bhatt2021spin,stanciu2020unexpected,becker2017ultrafast,stavrou1999magnetic}.

It has previously been observed that in transition-metal-Gd thin films, the Gd may segregate towards the surfaces~\cite{Kim2019} where oxidation can occur~\cite{bergeard2017correlation}. Aside from oxidation, a loss in Gd moment has also been reported for decreasing CoGd thickness in Ir/CoGd/Pt multilayers~\cite{streubel2018experimental}, suggesting a detrimental role of the interfaces with the transition-metal-Gd alloy. These material-specific phenomena, in addition to the more generic trend towards flux closure in thick films, add plausibility to distinct magnetic behaviors close to the sample surfaces compared to its bulk. 

\section{Conclusions}
\label{sec:conclusions}

We presented the first full-vectorial magnetic tomography based on Fourier transform holography achieving a resolution of $80$, $75$ and $60$\,nm in $m_x$, $m_y$ and $m_z$, respectively. To that end, we used a sample with two slits as holographic references which allowed us to probe all three components of the magnetization within the sample.

We acquired the magnetic projections by deconvoluting the object from these references. The recovered image is complex-valued and, in particular, its phase is proportional to the magnetic projection. Measuring the phase at the pre Fe $L_3$-edge also allows us for high contrast in soft x-rays even in a $800$\,nm-thick sample.

To validate our reconstruction method, we studied the effect of having a reduced angular range for tilting the sample and found that the missing wedge does not affect the recovering the out-of-plane magnetization nor the domain walls but it can fail recovering strong magnetic inhomogeneities or small domain caps.

To avoid reconstruction artifacts due to the missing information in the borders, we propose to use in the future patterned (finite) systems when utilizing FTH tomography with the dual-axis setup. For extended samples, the laminography setup represents a promising alternative since the information in the border of the disc is not lost. A future challenge will be to implement the aforementioned setup for FTH.

The resolution of the measurement is currently limited by the width of the reference and the sharpness of its ends, while the 3D reconstruction does not degrade it. It could in principle be significantly higher than demonstrated here, as 2D FTH images can be achieved down to 17\,nm resolution at $3d$ transition metal $L$ edges~\cite{turnbull2020tilted}.

Magnetic tomography by FTH can take advantage of the fairly simple FTH setups, which allow large and various sample environments. It could for instance be performed under applied magnetic field using a multi-coil rotatable magnetic field~\cite{popescu2019comet} opening up the study of the either static or even dynamic response of the 3D magnetic configuration to this stimulus.



\section*{Acknowledgments}
We acknowledge SOLEIL and the ESRF for providing synchrotron beamtime under project numbers 20201624 and MI-1384 respectively. We acknowledge the Agence Nationale de la Recherche for funding under project number ANR-19-CE42-0013-05 and the CNRS for the grant Emergence@INC2020. C. Donnelly acknowledges funding from the Max Planck Society Lise Meitner Excellence Program. The authors acknowledge Laurent Cagnon (Institut N\'{e}el) for the EDX measurements.

\appendix
\section{}

In a well conceived FTH experiment, the Fourier transform of the scattered intensity $g(\textbf{r})$ measured in the far field provides the convolution between the exit wave from the sample $\Psi(\textbf{r})$ and its inverse: $g(\textbf{r}) = \Psi(\textbf{r})\star\Psi(\textbf{-r})$. The exit wave can be considered as the sum over the exit wave from the object of interest $\Psi_O(\textbf{r})$ and the exit wave from the reference $\Psi_R(\textbf{r})$. A region of interest in $g(\textbf{r})$ provides one of the cross-terms between object and reference: $\Psi_O(\textbf{r})\star\Psi_R(\textbf{r})$. For the sake of simplicity, we assume in the following that the reference wave is a Dirac function, such that we consider the extracted term as the exit wave from the object $\Psi_O(\textbf{r})$. The HERALDO approach with an infinitely sharp slit, which is the one we use in this paper, yields the same result after the application of a linear filter \cite{guizar2007holography}.

The exit wave results from the propagation through the sample of the incident wave. Assuming an incident flat wave, the exit wave can be expressed as
\begin{equation}
\Psi(x,y) = \exp\left(\frac{2 \pi i}{\lambda} \int n \mathrm{d}z \right),
\label{eq:ewave}
\end{equation}
where $\lambda$ is the wavelength, $n$ is the optical index, the integration is along the beam axis $z$ and $x$ and $y$ are the transverse coordinates. 
The optical index includes a magnetic part, which will be detailed below. 

In many published works using FTH, the real part of $\Psi(x,y)$ is used, since it has shown to give good qualitative images for the magnetization~\cite{eisebitt2004lensless,streit2009magnetic,turnbull2020tilted,birch2020real,duckworth2011magnetic,duckworth2013holographic}. However, in order to perform tomography, a quantitative set of projections is needed. These are images that provide a quantity directly proportional to the magnetization of the sample. In that case, we notice that the real part actually consists in a mix between the absorption and refraction effects, both with magnetic components.
Therefore, we take the phase instead, which includes only refraction effects. The phase of $\Psi(x,y)$ is 
\begin{equation}
    \Phi(x,y) = \frac{2 \pi}{\lambda} \int n^\prime \mathrm{d}z,
\label{eq:phase}
\end{equation}
where $n^\prime$ is the real part of the optical index. Eq.~(\ref{eq:phase}) remains correct as long as the phase spans over less than 2$\pi$.

Next we will detail the magnetic dependence of the optical index and its circular dichroism. The optical index reads:
\begin{equation}
n = 1 - \frac{r_e \lambda^2}{2\pi} \rho f,
\label{eq:oindex}
\end{equation}
where $r_e$ is the classical electron radius, $\rho$ the density of scatterers and $f$ their atomic scattering factor. At an absorption edge of the scatterers, when the incident beam is circularly polarised, the atomic scattering factor can be written as
\begin{equation}
f = f_c \pm f_m \hat{m}\cdot\hat{k},
\label{eq:scatt}
\end{equation}
where $f_c$ corresponds to the electron density factor, $f_m$ to the dichroic scattering factor and $\hat{m}\cdot\hat{k}$ is the magnetization component along the beam direction~\cite{Hannon1988,van2008soft}.
$f_c$ and $f_m$ are resonant spectroscopic terms with generally both real and imaginary parts, \ie{}, $f_m=f'_m+i f''_m$. The sign of the magnetic term in Eq.~(\ref{eq:scatt}) changes with the helicity of the circular polarization. We point out that in the following, we will consider only the (resonant) scattering factors of iron. The contributions of Gd to magnetic scattering are negligible in our case, since we are measuring several hundreds of eV away from any absorption edge of Gd.

Combining Eqs.~(\ref{eq:phase}), (\ref{eq:oindex}) and (\ref{eq:scatt}), and assuming $\rho f_m$ constant (\ie{}, assuming the chemical homogeneity of the sample), we obtain the circular dichroism applied to the phase $\Phi(x,y)$ of the FTH reconstruction:
\begin{equation}
    \Delta\Phi(x,y) = -r_e \lambda\rho f_m^\prime \int \hat{m}\cdot\hat{k} \mathrm{d}z
    \label{eq:deltaphi}
\end{equation}
We see that the dichroic phase shift is proportional to the integrated projection onto the beam axis.

\begin{figure}[!htb]
     \includegraphics[width=\linewidth]{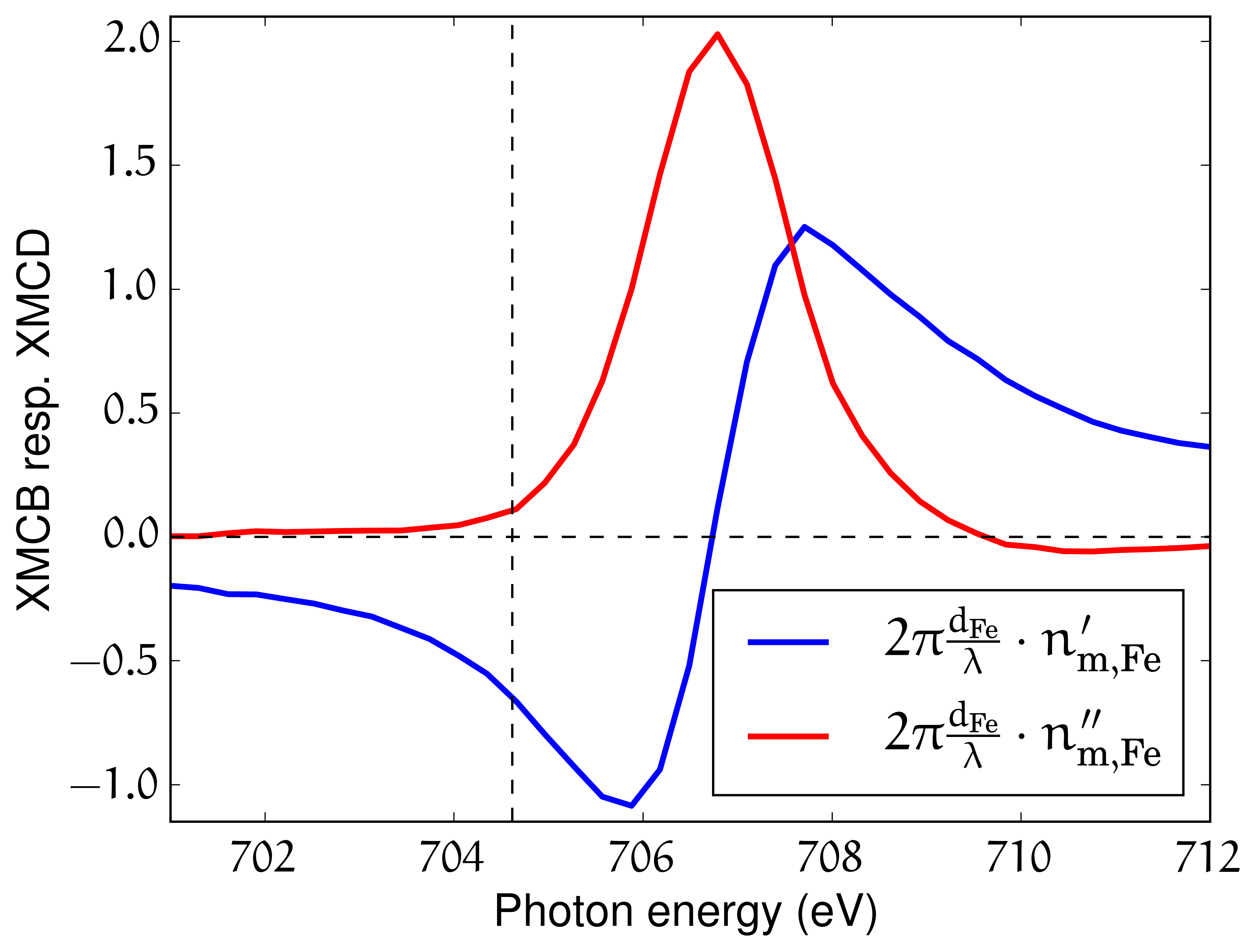}
     \caption{Calculated phase shift $\Delta\Phi=2\pi n'_{\mathrm{m,Fe}} d_\mathrm{Fe}/\lambda$ and absorbance $2\pi n''_{\mathrm{m,Fe}} d_\mathrm{Fe}/\lambda$ induced by the magnetization in the Fe/Gd multilayer for a positive photon helicity, assuming saturated magnetization along the beam. The total Fe thickness is $d_\mathrm{Fe}=$255 nm. The real $n'_{\mathrm{m,Fe}}$ and imaginary $n''_{\mathrm{m,Fe}}$ parts of the resonant magnetic contributions to the refractive index derive from the atomic scattering factor using $n_\mathrm{m,Fe}=-\frac{r_\mathrm{e}\lambda^2}{2\pi}\rho_\mathrm{Fe} f_\mathrm{m,Fe}$; the latter was taken from the measurements by Chen \emph{et al.}~\cite{chen1995_atScatFactors}. At the used photon energy, we find a phase shift of -0.649~rad, and an absorbance of 0.107.
     }
     \label{fig:magphaseshift}
\end{figure}

\begin{figure}[!htb]
 \includegraphics[width=\linewidth]{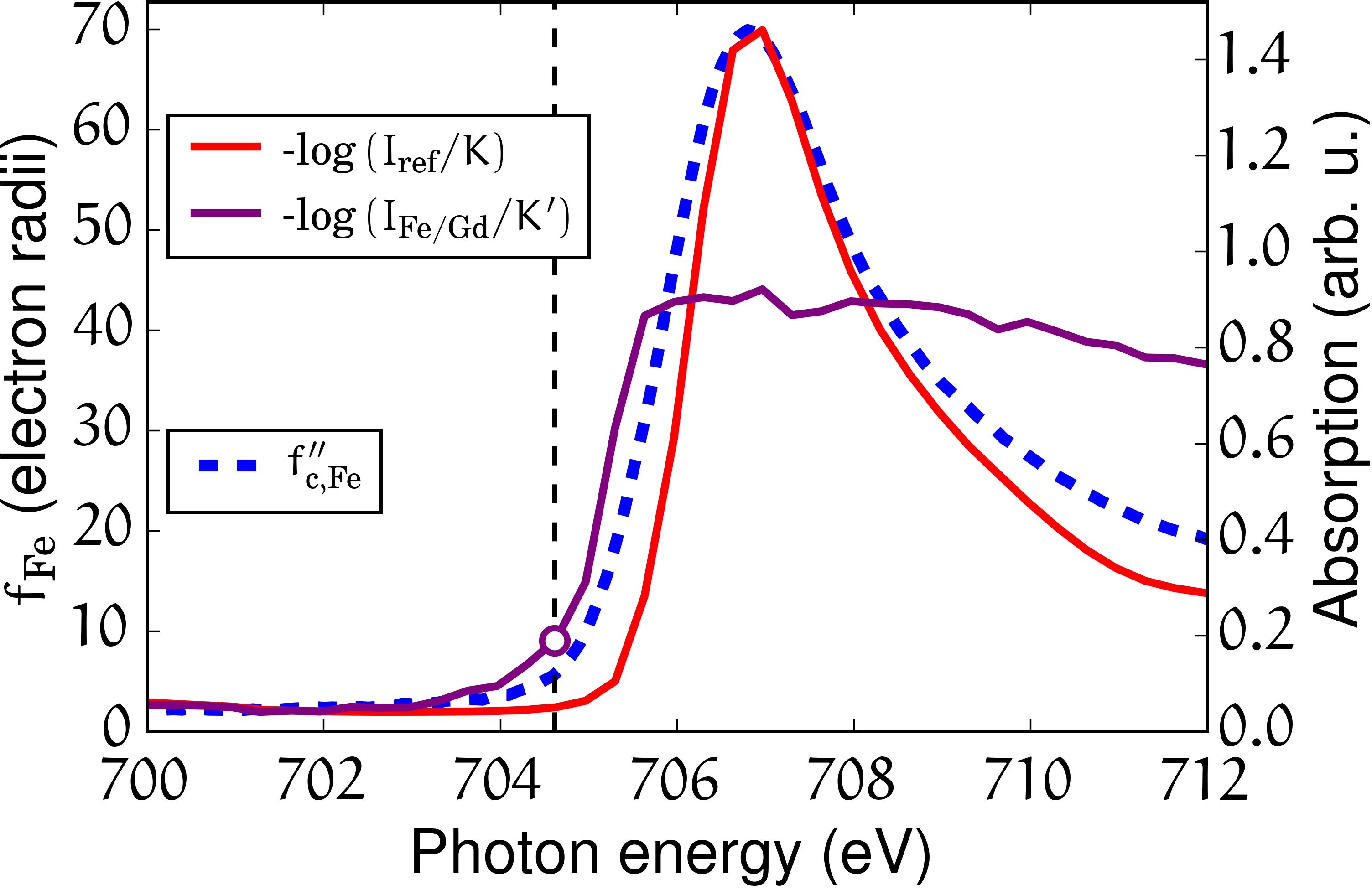}
 \caption{ Spectroscopic data  pertaining to Fe (dashed blue line) and transmission measurements on a reference Fe filter as well as our Fe/Gd multilayer (full lines, $K$ and $K'$ are arbitrary normalization constants). The imaginary part of the atomic scattering factor \cite{chen1995_atScatFactors} related to the charge ($f''_\mathrm{c,Fe}$, blue dashed line) and the intensity transmitted through a thin Fe film (red full curve) were used to remove the small energy offset ($\sim 1$\,eV) between the data obtained by Chen \etal{} and the measurements from SEXTANTS. The photon energy we used, $\sim 2.1$\,eV below the peak of absorption, is indicated by the vertical dashed line. It lies at the onset of the measured transmission through the multilayer, and yet corresponds to the best signal-to-noise ratio we obtained with FTH.}
 \label{fig:spectra}
\end{figure}

As mentioned above, Eq.~(\ref{eq:phase}) is valid a long as the phase shift spans over less than 2$\pi$, otherwise phase wraps will appear. If we assume the chemical homogeneity of the sample, the problem applies only to the dichroic part. According to Eq.~(\ref{eq:deltaphi}), in the case of saturated magnetization along the beam direction, the phase shift is
\begin{equation}
    \Delta\Phi = -r_e \lambda\rho f_m^\prime d
    \label{eq:deltaphimax}
\end{equation}
where $d$ is the thickness of the magnetic material. In the sample studied here, the total Fe thickness is around 255 nm, which corresponds to a phase shift around $\sim$0.6 rad at the energy of the measurement, well below the absorption edge (Figs.~\ref{fig:magphaseshift} and \ref{fig:spectra}). 

In contrast, the absorbance at the same energy is much lower, such that the absorption contrast would be very poor. At the peak of the magnetic absorbance, the phase contrast would vanish and the absorption contrast would be highest \cite{Donnelly2016}, but the dependence of the absorption contrast on the magnetization is only approximately linear, for a sufficiently optically thin sample.
The same holds for the real part, which is then the main reason why using it to calculate quantitatively the magnetization projection is only valid for thin samples.

\bibliography{vecfth.bib}

\end{document}